\documentclass[a4paper,fleqn]{cas-dc}

\usepackage[numbers]{natbib}

\def\tsc#1{\csdef{#1}{\textsc{\lowercase{#1}}\xspace}}
\tsc{WGM}
\tsc{QE}
\tsc{EP}
\tsc{PMS}
\tsc{BEC}
\tsc{DE}

\usepackage{hyperref}
\usepackage{tabularx}
\usepackage[utf8]{inputenc}
\usepackage{fancyhdr}
\usepackage{multirow}
\usepackage{amsmath,amsfonts,amssymb,amsthm}
\usepackage{enumerate} 
\everymath{\displaystyle}
\usepackage{caption}
\usepackage{tikz}
\usepackage{graphicx} 
\usepackage{bm}
\usepackage{sidecap}
\usepackage{subfig}
\usepackage{wrapfig} 
\usepackage{listings}
\theoremstyle{definition}
\usepackage{hyperref}
\usepackage{xcolor}
\usepackage[normalem]{ulem}
\usepackage{amssymb}
\usepackage{algorithm}
\usepackage[noend]{algpseudocode}
\usepackage{amsmath}
\usepackage[toc,page]{appendix}



\usepackage{booktabs}
\usepackage{multirow}
\usepackage{adjustbox}

\begin{document}
\let\WriteBookmarks\relax
\def\floatpagepagefraction{1}
\def\textpagefraction{.001}
\shorttitle{Real-time anomaly detection in base station testbeds}
\shortauthors{J. Witulska et~al.}

\title[mode=title]{Real-time anomaly detection in base station testbeds via scalable kernel density estimation framework}

\author[1]{Justyna Witulska}[orcid=0000-0003-4351-0783]
\cormark[1]
\ead{justyna.witulska@pwr.edu.pl}
\credit{Conceptualization, Methodology, Software, Writing -- Original Draft}

\affiliation[1]{organization={Faculty of Pure and Applied Mathematics and Hugo Steinhaus Center,
                Wrocław University of Science and Technology},
                city={Wrocław},
                postcode={50-370},
                country={Poland}}

\author[2]{Marcin Szczukiewicz}
\ead{marcin.szczukiewicz@nokia.com}

\author[2]{Artur Tabaka}
\ead{artur.tabaka@nokia.com}

\author[2]{Rafał Sarniak}
\ead{rafal.sarniak@nokia.com}

\affiliation[2]{organization={Nokia Solutions and Networks},
                city={Wrocław},
                postcode={54-130},
                country={Poland}}

\author[3]{Ireneusz Jabłoński}
\ead{ireneusz.jablonski@b-tu.de}

\affiliation[3]{organization={Faculty of MINT, Brandenburg University of Technology},
                city={Cottbus},
                postcode={03046},
                country={Germany}}

\author[1]{Agnieszka Wyłomańska}
\ead{agnieszka.wylomanska@pwr.edu.pl}

\author[2]{Dominik Dulas}
\ead{dominik.dulas@nokia.com}

\cortext[cor1]{Corresponding author}

\begin{abstract}
Large-scale testing infrastructures are critical for validating telecommunication systems, yet their growing complexity makes efficient resource utilization and anomaly detection increasingly challenging. In reservation-based testbed environments, errors in resource allocation or preparation often manifest as abrupt spikes or regime changes in time-based metrics. This paper proposes a scalable, unsupervised framework for real-time anomaly detection in such environments. We introduce CALM (Continuous Anomaly Localization for univariate and Multivariate data), a nonparametric method based on kernel density estimation and bootstrap-based thresholding, designed for anomaly detection at the individual testbed level. To address system-wide visibility, we further propose AggCALM, an aggregation framework that consolidates local anomaly signals across multiple testbeds to detect statistically significant global anomalies while mitigating alarm fatigue. The methodology is evaluated using simulated multivariate data and real-world data from a large-scale base station testing platform. Results demonstrate that the proposed framework enables timely, flexible, and accurate anomaly detection without requiring labeled data, supporting reliable operation of complex test environments. {Although the proposed methodology is presented within the context of a telecommunication testing labs, it can be effectively used to other applications, such as condition monitoring, where anomaly detection serves as a pivotal pre-processing step for diagnostic signals.}
\end{abstract}


\begin{keywords}
base station testbeds \sep anomaly detection \sep failures detection \sep non-Gaussian data \sep software testing
\end{keywords}

\maketitle

\section{Introduction}
\label{sec:intro}
\subsection{Problem description and motivation}

Complex systems are common in industries like IT, manufacturing, and healthcare \cite{barabasi2007architecture}. These systems involve interconnected components and dynamic, interleaving processes that result in the emergence of new properties and behaviors that are difficult to predict \cite{barabasi_Bursts}. For engineers responsible for monitoring and managing this complex infrastructure, controlling its temporal state is a challenge, especially in situations where these higher-order properties, emerging from simple interactions between many individual components, impair the operation of the entire system \cite{fortuna2022smart, panek2024automatic}. In this sense, these unpredictable states can be interpreted as failures and sources of inefficiency, or simply anomalies. 
Efficient resource allocation and anomaly detection are essential to keep operations running smoothly and reliably \cite{he2019intra_task_scheduling, hantom2022survey_Scheduling_Algorithms, stuber2023survey_TSN_Scheduling}. 

For managing complex systems, the initial phase of software and hardware testing of testbeds plays a crucial role. Testing has long served as both an acceptance criterion and a quality-assurance mechanism. Modern engineering practices have shortened the feedback loop between design, implementation, and verification through approaches such as continuous integration and continuous delivery/deployment \cite{panek_ASOC}, enabling earlier detection of issues and reducing correction costs. In telecommunication networks, testing is further complicated by the combination of custom software and purpose-built hardware such as radios and baseband units \cite{GSMAOpenRAN2020}, resulting in significantly higher complexity than in other domains \cite{Shahin2017CIReview}. 

Large-scale testing infrastructures are critical for validating telecommunication systems, yet their increasing scale and complexity make efficient resource utilization and anomaly detection increasingly difficult to manage. In testing environments, reservation systems are employed to manage shared resources, acting as queuing mechanisms that provide access, avoid bottlenecks, and ensure optimal utilization \cite{bharadwaj2020resource_reservaton_patent, ginthoer2022method_apparatus_reservation, djurayev2025application_query_theory, voudouris2017timing_anomaly_free}. Importantly, in reservation-based testbed environments, errors in resource allocation or preparation can easily go unnoticed, often manifesting only as abrupt spikes or regime changes in time‑based metrics. 

In 4G, 5G, and 6G mobile networks, Nokia's base station software testing platform exemplifies advanced resource management in such systems. It provides access to thousands of geographically distributed testbeds used to verify the functionality and performance of mobile base transceiver stations, where high operational costs make efficient utilization essential. Growing test demand and the need for parallel usage have rendered manual coordination impractical, leading to automated reservation and scheduling systems that enable continuous and conflict-free execution. As automation expanded to globally shared testbed pools, maintaining system health became increasingly complex, with centralized scheduling and continuous operation introducing new failure modes and reinforcing the need for automated anomaly detection. Reservation-based system was designed to streamline access and optimize resource usage. It ensures that the testbeds are used efficiently while maintaining high availability for testing needs. Furthermore, the reservation system generates time-based metrics that can be recorded and analyzed to detect non-standard behavior (anomalies), identify patterns, and derive operationally meaningful insights. Such anomalies may arise either from faults in the system or from changes in its underlying operational characteristics.

In general, software defects and errors can be introduced and may manifest at every stage of the software development life cycle. Mistakes in a programmer’s underlying logic may arise during implementation, while faults in the compiler or build system can lead to defective software distributions. Likewise, failures in the software environment or in third-party systems accessed through APIs can disrupt an otherwise well-designed application flow.

{The problem classification groups anomalies according to the dominant source of failure. IEEE Std 1044-2009 \cite{ieee_standard_class} provides a general framework for recording and analyzing software anomalies, but it does not define a precise or fixed taxonomy of defect types. Because such detailed categorical specifications are not defined in the standard, this study adopts the defect taxonomy used in \cite{hayes1997personal} -- which classifies defects into ten practical categories suitable for operational analysis.}

{In the case-study analyzed in this paper we have following categories: \textit{Environment, Interface, Build (Package), System, and Function}. The \textit{Environment} category refers to operational conditions external to the testbed infrastructure, such as maintenance windows or the temporary unavailability of supporting systems or other problems related. \textit{Interface} anomalies correspond to communication and integration issues with external systems, typically manifested as I/O or connectivity disruptions. The \textit{Build/Package} category includes failures related to dependency management or errors connected to version control. \textit{System} anomalies are associated with the runtime state of the infrastructure itself (configuration/memory etc.), whereas \textit{Function} refers to defects in the orchestration or control software logic (e.g., logic, pointers, loops, recursion). Finally, it is worth to note that effective monitoring should support not only fault detection, but also the classification needed to guide an appropriate remediation strategy}

A major challenge in systems - like Nokia's testing platform - is balancing the detection of production-disrupting problems with the prevention of alarm fatigue, where excessive notifications overwhelm operators and hinder their ability to respond effectively. This motivates practitioners to build the anomaly detection systems able prioritize alerts that are significant and actionable, while minimizing unnecessary notifications \cite{chasm}.

This paper tackles two main challenges: identifying real anomalies in individual environments and analyzing aggregated anomalies globally (across multiple testbeds) to find statistically significant deviations. By using widely available metrics from the reservation system, this approach ensures scalability and shows potential for reuse in other industries and smart infrastructure where resource optimization and anomaly detection are important.

\subsection{State of the Art, Research Gaps, Contributions}

This article focuses on real-time anomaly detection in base transceiver station testbeds. Historical data from reservation systems provides a baseline, but systems must distinguish between expected improvements and unexpected deviations. Statistical anomaly detection is effective for identifying events across measurable data, including unforeseen problems, but still requires separate investigation to determine root causes.

Existing research has explored anomaly detection methods, but scaling them to large, distributed environments like wireless telecommunication (e.g. Nokia's) testbeds remains challenging, as evolving software and hardware increase variability and the heterogeneity of testbed configurations further complicates anomaly identification. Prior approaches often focus on localized detection, overlooking global patterns and evolving baselines \cite{makris2020service_testbed, margery2014resources_testbed}.

The simplest methods used in industry, e.g. based on statically defined thresholds, are insufficient for dynamic environments. Statistical techniques \cite{Goldberg_2015} show promise but need customization to handle Nokia’s scale and diversity. {In \cite{marcin_cytowanie_ad}, a global-level method for identifying issues in test environments based on machine learning algorithms is presented; however, its reliance on Gaussian-based estimation reduces effectiveness for sparse, positively supported, and distinctly non-Gaussian (heavy-tailed) data distributions. Furthermore, the online framework assumes complete train-test coverage across all testbeds, which is impractical for the considered use case, where the reservation occurance frequency of an individual object depends on its topology and current utilization requirements.
}

In general, anomaly identification methods are used in many system monitoring applications. There is a wide range of methods for identifying anomalies in both univariate and multivariate data \cite{Raj2024ACS}, although in the latter case, most methods perform most effectively in an offline setting (i.e., not real-time). Anomaly detection methods can be categorized into the following groups: methods based on standard statistics, probabilistic methods, machine learning-based methods, hybrid techniques etc. \cite{samariya2023comprehensive, pyod_package, ghost}. Since many of these methods rely on extensive feature engineering, labeled data, or computationally intensive training, their practical deployment in real-time, large-scale environments is constrained \cite{habeeb2019real, samariya2023comprehensive}.

This work addresses these gaps by introducing a scalable, unsupervised framework for real-time anomaly detection tailored to reservation-based Nokia’s testbeds. We propose CALM, a nonparametric method for detecting anomalies at the individual testbed level using universally available metrics, and AggCALM, an aggregation mechanism that consolidates local anomaly signals to identify statistically significant system-wide events. The proposed framework improves global visibility, adapts to evolving baselines, and mitigates alarm fatigue through prioritized alerts, enabling reliable operation of complex, distributed testing environments.

{Although the proposed methodology is evaluated within the context of a specific testing lab problem, it possesses a universal character and is applicable to other domains where real-time anomaly detection remains a significant challenge. A prominent example of such an application is condition monitoring area, where the identification of anomalies in signals (e.g., vibrations) is a critical challenge, see e.g.  \cite{KUZIO2025113166}.}

The article is structured as follows. Section \ref{sec:methodology} provides a description of the proposed methodology for anomaly detection, called CALM and the aggregation framework called aggCALM. The simulation study is presented in Section \ref{sec:simulated_data}, while Section \ref{sec:real_data} provides a real data example of anomaly detection for telecommunication data. Finally, in Section \ref{sec:summary} we sum up all research.

\section{Methodology}
\label{sec:methodology}

This section introduces a novel methodology, denoted as \textbf{CALM} (\textbf{C}ontinuous \textbf{A}nomaly \textbf{L}ocalization for univariate and \textbf{M}ultivariate data), that facilitates real-time identification of anomalous observations in multivariate data. It uses the kernel density estimation (KDE) to obtain the probability density function (PDF) of training data (i.e., historical data without any anomalies). Then, the anomalies are identified by scoring the density estimation and comparing them against a threshold determined by bootstrap method. We applied a simplification of using multiple marginal densities estimators as input for scoring function instead of a multivariate density estimator (MDE), because classical MDE (such as histograms, kernel methods, and mixture models) struggle to scale beyond low dimensions because of the curse of dimensionality or limited nonparametric flexibility even with adaptive extensions \cite{peng2024nonparametric}. Although deep generative models have recently achieved impressive empirical performance in high-dimensional settings - their statistical guarantees remain to be a challenge \cite{peng2024nonparametric}. 

The simplification explained above means that the algorithm does not take into account the dependencies between individual components (dimensions), but focuses on the marginal properties. However, it aggregates anomaly information from a single dimension across multiple dimensions. In the real application discussed in this paper, the dependencies between variables may be neglected, because the analyzed variables concern independent operations performed on testbeds (resource allocation, environmental preparation, etc.). However, information from each variable provides us with greater insight into the phenomenon under study. Let $x_1, x_2, \ldots, x_N, \ (N \in \mathbb{N})$ denote a collection of independent and identically distributed observations drawn from a univariate distribution whose density function $f$ is unknown at an arbitrary point $x$. KDE is given as follows:

\begin{align}
 \mathcal{K}(x)={\frac {1}{Nh}}\sum _{i=1}^{N}K{\left({\frac {x-x_{i}}{h}}\right)},
\end{align}
where $K$ is a non-negative function named \textit{kernel}, and $h > 0$ is called \textit{bandwidth} (i.e., a smoothing parameter). For more details about KDE, see \cite{silverman2018density}.

Detailed information about the proposed methodology is provided in the subsections below. Initially, we present the technical formulation and operational principles of the CALM algorithm. Subsequently, we describe the general framework for global anomaly detection, wherein CALM constitutes the core component responsible for local anomaly scoring. The methodology is presented in multivariate scenario, however, it is also applicable for the univariate data.

\subsection{Theoretical background and description of the general idea}\label{math_problem}

Let us denote $d$-dimensional independent random vectors as $\mathbf{X} =\mathbf{X}_1, \mathbf{X}_2, \ldots, \mathbf{X}_N$ $(N \in \mathbb{N})$. We assume that the vectors are homogeneous within the segments corresponding to splitting $S_N^R$ defined as follows:
\begin{align*}
S_N^R = \{ (n_1, ..., n_{R-1}) \in \mathbb{N} ^ {R - 1} : 1 < n_1 < ... < n_{R-1} < N \}
\end{align*}
that refers to the following model:
  \begin{align}
\mathbf{X}_i \overset{d}{=} \begin{cases}
     \mathbf{X}^{(1)} \sim \mathbf{f}_1, \quad &\text{for }  1 \leq i < n_1,\\
     \mathbf{X}^{(2)} \sim \mathbf{f}_2, \quad &\text{for }  n_1 \leq i < n_2, \\
     ... \\
     \mathbf{X}^{(R)} \sim \mathbf{f}_{R}, \quad &\text{for }  n_{R-1} \leq i < N,
    \end{cases}
    \label{eq:assumption_multiple_cp}
\end{align}
where $n_1, ..., n_{R-2}, n_{R-1}$ are the change points and $\mathbf{f}_1, \mathbf{f}_2, \ldots, \mathbf{f}_{R-1}, \mathbf{f}_{R}$ represent PDFs of $d$-dimensional random vectors corresponding to the following segments: $[1, n_1), [n_1, n_2), \ldots$, $[n_{R-2}, n_{R-1}), \ldots [n_{R-1}, N]$, respectively. PDFs $\mathbf{f}_1, \mathbf{f}_2, \ldots, \mathbf{f}_{R-1}, \mathbf{f}_{R}$ may represent different distributions, or the same distributions but with different parameters (which indicate e.g. change in mean, scale, kurtosis, etc.).

The adopted definition of anomaly requires the introduction of additional notations. Let $X_i^{(j)}$ be a $j$-th component of $\mathbf{X}_i$ and let $\mathbf{x}_i$ be a realization of $\mathbf{X}_i$ with $j$-th component $x_i^{(j)}$. Let us denote $f_{i,j}^{\Theta}$ as a PDF of random variables ${X}^{(j)}_{i-\Theta}, {X}^{(j)}_{i-\Theta+1},$ $\ldots, {X}^{(j)}_{i-1}$. We introduce the \textit{scoring function} given by the formula:
\begin{equation}
    s(x; \Theta, j) := - \sqrt{(exp(f^{\Theta}_{i,j}(x))}, \ x \in \mathbb{R}.
    \label{eq:scoring_funct_theor}
\end{equation}
\textcolor{black}{The proposed \textit{scoring function} was selected based on an empirical analysis of real dataset. In this formulation, higher estimated density values produce more negative scores, while the square-root transformation reduces the severity of anomaly ranking compared with e.g.,  a logarithmic scoring function. A comparative evaluation of $s(x;\Theta,j)=- \sqrt{(exp(f^{\Theta}_{i,j}(x))}$ and $s(x;\Theta,j)= -log{(exp(f^{\Theta}_{i,j}(x))} = -f^{\Theta}_{i,j}(x)$ is provided in the Appendix. The accompanying GitHub\footnote{\textcolor{black}{The link will be provided after review.}} implementation allows users to select the preferred scoring function according to their application requirements; additional implementation details are also available in the Appendix.}

Let $\mathcal{K}^{\Theta}_{i,j}$ be the KDE estimated for $x^{(j)}_{i-\Theta}, x^{(j)}_{i-\Theta+1},$ $\ldots, x^{(j)}_{i-1}$. Then, the estimated scoring function for $j$-th component is given by:

\begin{equation}
    \tilde{s}(x; \Theta, j) := - \sqrt{(exp(\mathcal{K}^{\Theta}_{i,j}(x))}, \ x \in \mathbb{R},
    \label{eq:scoring_funct_est}
\end{equation}
and corresponding threshold $\tau_j$ is calculated as an empirical quantile of $\tilde{s}(x; \Theta, j)$ obtained in multiple replications (using the bootstrap method). For more details, see step 1 in the Sec. \ref{sec:proposed_approach}.

Based on scoring functions and thresholds calculated for each dimension $j \in \{1,...,d\}$ we evaluate whether the multivariate observation $\mathbf{x} \in \mathbb{R}^d$ is an anomaly or not. The criterion for identification of anomalies in multidimensional scenario is as follows:
\begin{align}
         L(\mathbf{x}) =
    \begin{cases}
    1, & \text{if for any } j: \tilde{s}(x^{(j)}; \Theta, j) > \tau_j, \\
    0, & \text{otherwise,}
    \end{cases}
    \label{agg}
    \end{align}
where $x^{(j)}$ is denoted as single $j$-th component of $\mathbf{x}$. Based on $L(\mathbf{x})$ the algorithm determines whether the multidimensional observation $\mathbf{x}$ constitutes an anomaly ($L(\mathbf{x})=1$) or not ($L(\mathbf{x})=0$).

\subsection{Description of the procedure for anomaly detection}
\label{sec:proposed_approach}

The proposed CALM method enables the real-time detection of anomalies in multivariate data. The possible delays arise only from the computation of the scoring function applied in the algorithm. The CALM is a non-parametric method that does not require any prior knowledge of the distribution of the input data. The algorithm requires the provision of training input data used for the estimation of KDE and the determination of threshold values based on the scoring functions \eqref{eq:scoring_funct_est}. These thresholds ($\tau_j, \ j \in \{1,...,d\}$) are used in the derived decision criterion given by \eqref{agg}.

CALM employs the following input parameters: (i) the number of bootstrap resamples $N_r$ utilized for dynamic threshold estimation, (ii) the bootstrap quantile level $q_b$ applied during the threshold estimation procedure, and (iii) the sample fraction $\phi$ serving as an auxiliary parameter in the bootstrap process. Let us denote $y_{k}^{(j)}$ as $j$-th component of $d$-dimensional vector $\mathbf{y}_k$. A step-by-step description of CALM is presented below:

\begin{enumerate}
\item Training phase
    \begin{enumerate}
    \item Load the training data without anomalies $\mathbf{y}_1,...,\mathbf{y}_N \in \mathbb{R}^d$. \\
    \item For each $j \in \{1,...,d\}$:
    \begin{itemize}
    \item Fit base KDE ($\mathcal{K}_j$) for $y^{(j)}$,
    \item Split the training set $y^{(j)}$ into two equal subsets: $Y_1$ and $Y_2$. For $i$ in $\{1,..., N_r\}$:
    \begin{itemize}
        \item Select a random subset from $Y_1$ and calculate KDE from it: $\tilde{\mathcal{K}}_j$.
        \item Select a random subset from $Y_2$ and for each observation $y_m^{(j)} \in Y_2$ calculate:    
        \[
        \tilde{s}(y_m^{(j)}) = -\sqrt{\exp(\tilde{\mathcal{K}}_j(y_m^{(j)})).}
    \]
    \end{itemize}
    \end{itemize}
    \item From all obtained $\tilde{s}(y_m^{(j)})$, calculate the empirical quantile at the level $q_b$. The value obtained is assigned as the  threshold $\tau_j$. Once the thresholds are established,  we can use them in prediction phase in which for each ’new observation’ (i.e., each observation which
was collected after the training phase) the score is evaluated.
\end{enumerate}
\item Prediction phase
    \begin{enumerate}
    \item Load new data dump $\mathbf{x}_1,...,\mathbf{x}_M \in \mathbb{R}^d$.
    For each $m \in \{1,...,M\}$:
    \begin{itemize}
\item For each $j \in \{1,...,d\}$ compute anomaly score:
    \[
        \tilde{s}(x_m^{(j)})= -\sqrt{\exp(\mathcal{K}_j(x_m^{(j)}))}.
    \]
    \item Combine feature-wise labels:
    \[
    L(\mathbf{x}_m) =
    \begin{cases}
    1, & \text{if for any } j:  \tilde{s}(x_m^{(j)}) > \tau_j, \\
    0, & \text{otherwise.}
    \end{cases}
    \]
    \end{itemize}
\end{enumerate}
\end{enumerate}

Importantly, the CALM can incorporate a retraining mechanism based on updated knowledge of the data, as the underlying data distribution (or its parameters, such as mean and variance) may change over time, as discussed above. Therefore, the retraining process aims to preserve the homogeneity of the data with respect to the underlying (non-anomalous) distribution and its parameters.

In the following part, we present the pseudocodes that correspond to the proposed procedure. Algorithm \ref{alg:train} refers to training phase and Algorithm \ref{alg:pred} refers to prediction phase. 


\begin{algorithm}
\caption{CALM (Training Phase)}
\label{alg:train}
\begin{algorithmic}[1]

\Require Training data matrix $\mathbf{Y} \in \mathbb{R}^{n \times d}$, 
number of bootstrap resamples $N_r$, bootstrap quantile $q_b$, sample fraction $\phi$

\Ensure KDE models $\{\mathcal{K}_j\}_{j=1}^d$, thresholds $\{\tau_j\}_{j=1}^d$

\State Initialize model list $\mathcal{K} \gets \emptyset$, threshold list $\mathcal{T} \gets \emptyset$

\For{$j = 1$ to $d$}

    \State Extract feature column $y_{\text{train}} \gets \mathbf{Y}_{:,j}$

    \State Fit base KDE: $\mathcal{K}_j \gets \text{KDE}(y_{\text{train}})$

    \State Initialize bootstrap score container $\mathcal{B}_j \gets \emptyset$

    \For{$r = 1$ to $N_r$}

        \State Draw resample $x_1', x_2'$ from $y_{\text{train}}$ of size $\phi|y_{\text{train}}|$

        \State Fit temporary KDE: $\mathcal{K}_r \gets \text{KDE}(x_1')$

        \For{each $v \in x_2'$}

            \State Compute score:
            \[
                s_j = -\sqrt{\exp(\mathcal{K}_r(v))}
            \]

            \State Append $s_j$ to $\mathcal{B}_j$

        \EndFor
    \EndFor

    \State Compute threshold $\tau_j \gets \text{Quantile}(\mathcal{B}_j, q_b)$

    \State Append $(\mathcal{K}_j, \tau_j)$ to $(\mathcal{K}, \mathcal{T})$

\EndFor

\State \Return $\{\mathcal{K}_j\}, \{\tau_j\}$

\end{algorithmic}
\end{algorithm}

\begin{algorithm}
\caption{CALM (Prediction Phase)}
\label{alg:pred}
\begin{algorithmic}[1]

\Require Observation data matrix $\mathbf{X} \in \mathbb{R}^{n \times d}$

\For{$j = 1$ to $d$}

    \State Extract feature: $x_j \gets \mathbf{X}_{:,j}$

    \State Compute anomaly score:
    \[
        s_j = -\sqrt{\exp(\mathcal{K}_j(x_j))}
    \]

\EndFor

\State Combine feature-wise decisions:
\[
L(\mathbf{x}) =
\begin{cases}
1, & \text{if } s_j > \tau_j \text{ for any } j, \\
0, & \text{otherwise}.
\end{cases}
\]

\State \textbf{Output:} Label $L(\mathbf{x})$ for each sample.

\end{algorithmic}
\end{algorithm}

\subsection{Framework for detecting global anomalies} 
\label{sec:aggregation_framework}
In this part, we describe the framework for detecting global anomalies (AggCALM), i.e., the procedure that performs global anomaly detection across multiple testbeds based on previously trained CALM models. The AggCALM requires the following input parameters: (i) the aggregation window $\omega$ used to collect and evaluate recent anomaly detections, (ii) the alarm threshold $T_a$ representing the minimum proportion of testbeds required to trigger a global alert, and (iii) the retraining interval $t_r$ defining the time period after which all CALM models are retrained to maintain detection accuracy under data drift. A step-by-step description of the proposed procedure is as follows.
For each variable $j$ and each testbed $k$, CALM identifies local anomalies using the most recent data dump. The algorithm then aggregates the detected anomalies over an observation window $\omega$ and computes the number of affected testbeds, denoted as $\tilde{T}_a$. If the proportion of anomalous testbeds exceeds the alarm threshold $T_a$, a global anomaly alert is raised for the corresponding variable. The entire process is repeated continuously for each new data dump. When the elapsed time exceeds the retraining interval $t_r$, all CALM models are retrained for the respective variables and testbeds to ensure model adaptability to data drift. Algorithm \ref{alg:global_ad} corresponds to the described procedure.
 
\begin{algorithm}
\caption{Global anomalies detection (AggCALM)}
\label{alg:global_ad}
\begin{algorithmic}[1]

\Require 
$d$-dimensional data streams from $K$ testbeds, 
retraining interval $t_r$, 
aggregation window $\omega$, 
and alarm threshold $T_a$ (percentage of affected testbeds)

\State Each CALM model has been pre-trained separately for every testbed and each variable $j=1,\ldots,d$ using one-dimensional historical data.

\For{each new data dump}
    \For{$j = 1$ to $d$}
        \For{$k = 1$ to $K$}
            \State Detect anomalies for testbed $k$ and variable $j$ using the corresponding pre-trained CALM model.
        \EndFor

        \State Determine which testbeds show anomalies within the last aggregation window $\omega$.
        \State Let $\tilde{T}_a$ be the number of such testbeds.

        \If{$\tilde{T}_a > T_a$}
            \State Raise a global anomaly alert for variable $j$.
        \EndIf
    \EndFor

    \State Continue processing incoming data dumps.

    \If{elapsed time $> t_r$}
        \State Retrain all CALM models for all testbeds.
    \EndIf
\EndFor

\end{algorithmic}
\end{algorithm}

\section{Assessment of the proposed procedure for multivariate simulated data}
\label{sec:simulated_data}

To evaluate the performance of our anomaly detection methods under controlled conditions, we designed a flexible simulation framework that generates realistic time-series data for multiple testbeds. The generator incorporates controlled noise, probabilistic anomalies, and variable base distributions. Below, we describe the simulation setup and its key components.

The adopted model describes a contaminated stochastic process in which deviations from the uncontaminated multivariate distribution occur through additive, directionally consistent perturbations affecting individual coordinates with probability $p$.

Let $\mathbf{X}_1, \mathbf{X}_2, \ldots, \mathbf{X}_N$ be a sequence of $d$-dimensional random vectors.  
A random vector $\mathbf{X}_t \in \mathbb{R}^d$ follows a multivariate stochastic process with contamination if it can be represented as
\begin{align}
\mathbf{X}_t = \mathbf{Y}_t + \mathbf{C}_t, \quad t = 1, \ldots, N,
\label{eq:cont_d_process}
\end{align}
where $\mathbf{Y}_t$ denotes uncontaminated time series and $\mathbf{C}_t$ represents a contamination process.

The uncontaminated component $\mathbf{Y}_t$ represents an arbitrary $d$-dimensional stochastic process. In this study, we analyze it as a multivariate Gaussian distribution or a VARMA model (vector ARMA), see, Sec. \ref{sec:methodology_gauss_contamination} and  Sec. \ref{sec:methodology_varmax}, respectively. However, this framework could be further extended for other models.

The contamination component $\mathbf{C}_t$ is defined element-wise as
$\mathbf{C}_t = \mathbf{M}_t \odot \mathbf{B}_t \odot \mathrm{sgn}(\mathbf{Y}_t), $
where
\begin{itemize}
    \item $\mathbf{M}_t \in \mathbb{R}^d$ is a random magnitude vector with independent components distributed uniformly on the interval $[a,b]$, i.e.
 $M_{t,j} \sim \mathcal{U}(a,b), \quad j = 1, \ldots, d,$ where $M_{t,j}$ is a $j$-th component of $\mathbf{M}_t$,
    \item $\mathbf{B}_t \in \{0,1\}^d$ is a Bernoulli contamination indicator vector with $B_{t,j} \sim \mathrm{Bernoulli}(p),$ where $B_{t,j}$ is a $j$-th component of $\mathbf{B}_t$, and $p$ is a contamination probability,
    \item $\odot$ denotes the Hadamard (element-wise) product,
    \item $\mathrm{sgn}(\cdot)$ is the element-wise sign function.
\end{itemize}

\noindent
We assume, an observation $\mathbf{X}_t$ is \textit{contaminated} if at least one of its components is affected, i.e., if ${\displaystyle{\sum_{j=1}^{d} B_{t,j} > 0}}$.
\subsection{Multivariate Gaussian distribution with contamination}
\label{sec:methodology_gauss_contamination}

Let $\mathbf{X}_1, \mathbf{X}_2, \ldots, \mathbf{X}_N$ be a sequence of $d$-dimensional random vectors.  
A random vector $\mathbf{X}_t \in \mathbb{R}^d$ follows a multivariate Gaussian distribution with contamination if it can be represented as \eqref{eq:cont_d_process}, where the uncontaminated component $\mathbf{Y}_t$ is assumed to follow a $d$-dimensional Gaussian distribution
\begin{align}
\mathbf{Y}_t \sim \mathcal{N}_d(\bm{\mu}, \bm{\Sigma}),
\label{eq:gauss_clean}
\end{align}
where  $\bm{\mu} \in \mathbb{R}^d$ is the mean vector, $\bm{\Sigma} \in \mathbb{R}^{d \times d}$ is the positive definite covariance matrix.

\subsection{VARMA model with contamination}
\label{sec:methodology_varmax}

Let $\mathbf{X}_1, \mathbf{X}_2, \ldots, \mathbf{X}_N$ be a sequence of $d$-dimensional random vectors.  
A random vector $\mathbf{X}_t \in \mathbb{R}^d$ follows a \textit{VARMA model with contamination} if it can be represented as \eqref{eq:cont_d_process}, where the uncontaminated component $\mathbf{Y}_t$ is a \textit{d-}dimensional ARMA(p,q) process defined as follows

\begin{align}
\Phi(F) \bm{Y_t} = \Theta(F)\bm{Z_t},
\label{eq:mARMA}
\end{align}
\noindent
where $Y_t$ is a \textbf{stationary solution} of difference equations  \eqref{eq:mARMA} and
\begin{itemize}
    \item $\Phi(z) := I - \Phi_1 z - ... - \Phi_p z^p,$ where $\Phi_1, ..., \Phi_p$ are $d\times d$ matrices,
    \item $\Theta(z) := I - \Theta_1 z - ... - \Theta_q z^q,$ where $\Theta_1, ..., \Theta_q$ are $d\times d$ matrices.
    \end{itemize}

In the above, we have the following notions:  I is $d\times d$ identity matrix, F is the backward shift operator, $\bm{Z_t}$ is multivariate white noise sequence. For more details about VARMA model, definition of multivariate white noise, and methods for estimation of parameters, see \cite{brockwell1991time}.

\subsection{Evaluating CALM over baselines}
\label{sec:baselines_sim_comparison}
In this part, we conduct a comparative analysis of the quality of classifying anomalous points as anomalies by comparing the proposed CALM method with approaches well established in the literature, namely: Unsupervised Outlier Detection Using Empirical Cumulative Distribution Functions (ECOD), Isolation Forest (IFOREST), Local Outlier Factor (LOF), Minimum Covariance Determinant Method (MCD), and One-Class Support Vector Machines (OCSVM) \cite{pyod_package}. These methods were selected because they represent classical approaches to anomaly detection and cover different classes of algorithms.

\textcolor{black}{All experiments were conducted on a workstation equipped with an \textit{Intel Xeon E5-2680 v4} processor with \textit{247 GB of RAM}, running \textit{Ubuntu 22.04.4 LTS} with Linux kernel version \textit{6.12.31-talos}. No GPU acceleration was used.}

We consider three simulation scenarios. In the first scenario, the data are generated according to the model~\eqref{eq:cont_d_process} with an uncontaminated baseline given by~\eqref{eq:gauss_clean}. In the second scenario, the data follow the model~\eqref{eq:cont_d_process} with an uncontaminated baseline defined by the VARMA process~\eqref{eq:mARMA} with Gaussian noise. The model order is set to $p=2$, $q=0$, there is no dependence between individual components, and the corresponding coefficients are equal to $0.2$ and $-0.3$. In the third scenario, the data are again generated from the model~\eqref{eq:cont_d_process} with an uncontaminated baseline given by~\eqref{eq:mARMA} with Gaussian noise, but the model order is $p=1$, $q=1$. The coefficients of the model are set to $0.3$ and $0.2$. The innovation covariance matrix is specified such that the first component is correlated with all remaining components, with correlation coefficient $\rho = 0.99 \cdot (1/(d-1)^2)$ to ensure positive definiteness, while the remaining components are mutually independent.

In all three scenarios, the length of the time series is fixed to $N=1000$, the noise range parameters are set to $a=5$ and $b=10$, and the contamination probability is equal to $0.05$. The parameters of the CALM method used in the simulations are $N_r = 1000$, $q_b = 0.95$, $\phi = 0.75$. 

The training data were prepared using offline anomaly detection methods to ensure that only non-anomalous samples are included in the thresholds determination process for each data dimension. Let us denote $y_{k}^{(j)}$ as $j$-th component of $d$-dimensional vector $\mathbf{y}_k$. The training data $y_{1}^{(j)}, y_{2}^{(j)}, \ldots, y_{N}^{(j)}$ are first processed using offline anomaly detection methods to ensure that only non-anomalous samples are included in the thresholds determination process for each data dimension $j$. The procedure uses the input parameters: $(q_{\text{low}}, q_{\text{high}})$ that are selected quantile bounds. The training data preprocessing for each dimension $j$ looks as follows:  

\begin{itemize}
\item  Calculate the sample mean ($\mu_j$), standard deviation ($\sigma_j$), and quantiles at levels $q_{low}$ and $q_{high}$ (mark them as $q_{1}$ and $q_{2}$, respectively) based on $y_{1}^{(j)}, y_{2}^{(j)}, \ldots, y_{N}^{(j)}.$
    \item  $y^{(j)}_{\text{train}} = \{y_k^{(j)}: \min\big(\text{q}_\text{1},\, \mu_j - 2\sigma_j\big) < y_k^{(j)} < \max\big(\text{q}_\text{2},\, \mu_j + 2\sigma_j\big), k \in \{1,...,N\}\}$.
\end{itemize} The parameters for training-set creation (using in this section) are as follows: $q_1 = 0.05$.

\begin{table*}[!t]
\centering
\caption{Precision for contamination probability $0.05$ (mean and standard deviation).}
\label{tab:precision_005}
\begin{adjustbox}{width=0.6\textwidth}
\begin{tabular}{cccccccc}
\toprule
\textbf{Sim.} & \textbf{Dim.} & \textbf{CALM} & \textbf{ECOD} & \textbf{IFOREST} & \textbf{LOF} & \textbf{MCD} & \textbf{OCSVM} \\
\midrule
\multirow{5}{*}{1}
& 2  & 0.764 (0.19) & 0.836 (0.08) & 0.803 (0.08) & 0.768 (0.10) & 0.910 (0.05) & 0.494 (0.05) \\
& 3  & 0.772 (0.19) & 0.870 (0.09) & 0.804 (0.06) & 0.775 (0.08) & 0.975 (0.02) & 0.878 (0.04) \\
& 5  & 0.784 (0.17) & 0.868 (0.10) & 0.994 (0.02) & 0.978 (0.03) & 0.998 (0.00) & 0.983 (0.02) \\
& 10 & 0.813 (0.15) & 0.968 (0.07) & 1.000 (0.00) & 1.000 (0.00) & 1.000 (0.00) & 0.885 (0.32) \\
& 25 & 0.890 (0.09) & 0.989 (0.07) & 1.000 (0.00) & 1.000 (0.00) & 1.000 (0.00) & 0.960 (0.06) \\
\midrule
\multirow{5}{*}{2}
& 2  & 0.775 (0.20) & 0.793 (0.12) & 0.801 (0.08) & 0.766 (0.10) & 0.890 (0.05) & 0.499 (0.05) \\
& 3  & 0.771 (0.19) & 0.859 (0.10) & 0.805 (0.07) & 0.771 (0.08) & 0.975 (0.03) & 0.808 (0.05) \\
& 5  & 0.780 (0.18) & 0.913 (0.09) & 0.812 (0.05) & 0.802 (0.06) & 0.997 (0.01) & 0.979 (0.03) \\
& 10 & 0.814 (0.15) & 0.964 (0.07) & 0.837 (0.03) & 0.843 (0.03) & 1.000 (0.00) & 0.759 (0.43) \\
& 25 & 0.889 (0.09) & 0.998 (0.02) & --            & --            & 1.000 (0.00) & 0.986 (0.03) \\
\midrule
\multirow{5}{*}{3}
& 2  & 0.826 (0.15) & 0.506 (0.11) & 0.857 (0.06) & 0.824 (0.10) & 0.920 (0.06) & 0.529 (0.06) \\
& 3  & 0.795 (0.17) & 0.714 (0.13) & 0.825 (0.06) & 0.776 (0.08) & 0.983 (0.02) & 0.659 (0.06) \\
& 5  & 0.792 (0.17) & 0.877 (0.10) & 0.974 (0.03) & 0.957 (0.05) & 0.999 (0.00) & 0.954 (0.03) \\
& 10 & 0.811 (0.14) & 0.937 (0.10) & 0.999 (0.00) & 1.000 (0.00) & 1.000 (0.00) & 0.820 (0.39) \\
& 25 & 0.887 (0.09) & 0.992 (0.04) & 1.000 (0.00) & 1.000 (0.00) & 1.000 (0.00) & 0.978 (0.04) \\
\bottomrule
\end{tabular}
\end{adjustbox}
\end{table*}

\begin{table*}[!t]
\centering
\caption{Recall for contamination probability $0.05$ (mean and standard deviation).}
\label{tab:recall_005}
\begin{adjustbox}{width=0.6\textwidth}
\begin{tabular}{cccccccc}
\toprule
\textbf{Sim.} & \textbf{Dim.} & \textbf{CALM} & \textbf{ECOD} & \textbf{IFOREST} & \textbf{LOF} & \textbf{MCD} & \textbf{OCSVM} \\
\midrule
\multirow{5}{*}{1}
& 2  & 0.905 (0.16) & 0.367 (0.08) & 1.000 (0.00) & 0.949 (0.09) & 0.999 (0.01) & 0.999 (0.01) \\
& 3  & 0.899 (0.15) & 0.174 (0.05) & 1.000 (0.00) & 0.955 (0.07) & 0.991 (0.01) & 0.873 (0.06) \\
& 5  & 0.903 (0.14) & 0.090 (0.03) & 0.895 (0.07) & 0.847 (0.08) & 0.926 (0.03) & 0.286 (0.06) \\
& 10 & 0.914 (0.13) & 0.039 (0.01) & 0.507 (0.05) & 0.501 (0.06) & 0.511 (0.07) & 0.017 (0.01) \\
& 25 & 0.939 (0.09) & 0.020 (0.01) & 0.276 (0.03) & 0.276 (0.03) & 0.076 (0.02) & 0.040 (0.02) \\
\midrule
\multirow{5}{*}{2}
& 2  & 0.901 (0.17) & 0.352 (0.09) & 0.998 (0.01) & 0.946 (0.10) & 0.998 (0.01) & 0.999 (0.01) \\
& 3  & 0.896 (0.16) & 0.181 (0.06) & 0.999 (0.01) & 0.952 (0.08) & 0.989 (0.02) & 0.821 (0.07) \\
& 5  & 0.902 (0.15) & 0.102 (0.04) & 0.903 (0.06) & 0.858 (0.07) & 0.931 (0.03) & 0.301 (0.07) \\
& 10 & 0.913 (0.13) & 0.045 (0.02) & 0.521 (0.06) & 0.513 (0.07) & 0.523 (0.08) & 0.019 (0.01) \\
& 25 & 0.938 (0.10) & 0.023 (0.01) & --            & --            & 0.082 (0.02) & 0.042 (0.02) \\
\midrule
\multirow{5}{*}{3}
& 2  & 0.918 (0.14) & 0.402 (0.10) & 1.000 (0.00) & 0.961 (0.08) & 0.999 (0.01) & 0.999 (0.01) \\
& 3  & 0.905 (0.15) & 0.221 (0.07) & 1.000 (0.00) & 0.957 (0.07) & 0.993 (0.01) & 0.881 (0.06) \\
& 5  & 0.907 (0.14) & 0.121 (0.04) & 0.921 (0.06) & 0.879 (0.07) & 0.942 (0.03) & 0.334 (0.07) \\
& 10 & 0.916 (0.12) & 0.052 (0.02) & 0.548 (0.06) & 0.539 (0.07) & 0.547 (0.08) & 0.021 (0.01) \\
& 25 & 0.941 (0.09) & 0.026 (0.01) & 0.291 (0.03) & 0.291 (0.03) & 0.089 (0.02) & 0.045 (0.02) \\
\bottomrule
\end{tabular}
\end{adjustbox}
\end{table*}

\begin{table*}[!t]
\centering
\caption{F1-score for contamination probability $0.05$ (mean and standard deviation).}
\label{tab:f1_005}
\begin{adjustbox}{width=0.6\textwidth}
\begin{tabular}{cccccccc}
\toprule
\textbf{Sim.} & \textbf{Dim.} & \textbf{CALM} & \textbf{ECOD} & \textbf{IFOREST} & \textbf{LOF} & \textbf{MCD} & \textbf{OCSVM} \\
\midrule
\multirow{5}{*}{1}
& 2  & 0.795 (0.11) & 0.504 (0.09) & 0.888 (0.05) & 0.846 (0.08) & 0.951 (0.03) & 0.660 (0.05) \\
& 3  & 0.799 (0.10) & 0.287 (0.07) & 0.890 (0.04) & 0.854 (0.07) & 0.983 (0.01) & 0.874 (0.04) \\
& 5  & 0.812 (0.09) & 0.161 (0.05) & 0.940 (0.04) & 0.905 (0.05) & 0.960 (0.02) & 0.440 (0.08) \\
& 10 & 0.840 (0.07) & 0.074 (0.03) & 0.671 (0.05) & 0.665 (0.06) & 0.674 (0.06) & 0.033 (0.03) \\
& 25 & 0.906 (0.03) & 0.040 (0.02) & 0.432 (0.04) & 0.432 (0.04) & 0.140 (0.03) & 0.077 (0.03) \\
\midrule
\multirow{5}{*}{2}
& 2  & 0.794 (0.12) & 0.491 (0.10) & 0.887 (0.05) & 0.842 (0.09) & 0.947 (0.03) & 0.665 (0.06) \\
& 3  & 0.796 (0.11) & 0.302 (0.08) & 0.889 (0.05) & 0.849 (0.07) & 0.981 (0.02) & 0.814 (0.05) \\
& 5  & 0.810 (0.10) & 0.182 (0.05) & 0.940 (0.04) & 0.902 (0.05) & 0.963 (0.02) & 0.454 (0.08) \\
& 10 & 0.839 (0.08) & 0.081 (0.03) & 0.681 (0.05) & 0.674 (0.06) & 0.683 (0.06) & 0.035 (0.03) \\
& 25 & 0.905 (0.04) & 0.044 (0.02) & --            & --            & 0.150 (0.03) & 0.081 (0.03) \\
\midrule
\multirow{5}{*}{3}
& 2  & 0.815 (0.10) & 0.541 (0.09) & 0.923 (0.04) & 0.885 (0.07) & 0.958 (0.03) & 0.682 (0.06) \\
& 3  & 0.804 (0.10) & 0.343 (0.08) & 0.908 (0.04) & 0.866 (0.07) & 0.988 (0.01) & 0.752 (0.05) \\
& 5  & 0.814 (0.09) & 0.214 (0.06) & 0.946 (0.04) & 0.918 (0.05) & 0.969 (0.02) & 0.498 (0.08) \\
& 10 & 0.842 (0.07) & 0.091 (0.03) & 0.706 (0.05) & 0.699 (0.06) & 0.705 (0.06) & 0.038 (0.03) \\
& 25 & 0.908 (0.04) & 0.049 (0.02) & 0.445 (0.04) & 0.445 (0.04) & 0.162 (0.03) & 0.085 (0.03) \\
\bottomrule
\end{tabular}
\end{adjustbox}
\end{table*}

\begin{table*}[!t]
\centering
\caption{Accuracy for contamination probability $0.05$ (mean and standard deviation).}
\label{tab:accuracy_005}
\begin{adjustbox}{width=0.6\textwidth}
\begin{tabular}{cccccccc}
\toprule
\textbf{Sim.} & \textbf{Dim.} & \textbf{CALM} & \textbf{ECOD} & \textbf{IFOREST} & \textbf{LOF} & \textbf{MCD} & \textbf{OCSVM} \\
\midrule
\multirow{5}{*}{1}
& 2  & 0.952 (0.03) & 0.931 (0.01) & 0.975 (0.01) & 0.966 (0.02) & 0.990 (0.00) & 0.899 (0.01) \\
& 3  & 0.932 (0.04) & 0.880 (0.02) & 0.964 (0.01) & 0.953 (0.02) & 0.995 (0.00) & 0.964 (0.01) \\
& 5  & 0.900 (0.06) & 0.789 (0.02) & 0.975 (0.01) & 0.961 (0.02) & 0.982 (0.01) & 0.836 (0.02) \\
& 10 & 0.855 (0.08) & 0.615 (0.02) & 0.803 (0.02) & 0.799 (0.03) & 0.802 (0.04) & 0.605 (0.02) \\
& 25 & 0.859 (0.05) & 0.291 (0.02) & 0.477 (0.03) & 0.477 (0.03) & 0.331 (0.03) & 0.305 (0.02) \\
\midrule
\multirow{5}{*}{2}
& 2  & 0.953 (0.03) & 0.930 (0.02) & 0.975 (0.01) & 0.963 (0.02) & 0.988 (0.01) & 0.904 (0.01) \\
& 3  & 0.931 (0.04) & 0.881 (0.02) & 0.964 (0.02) & 0.952 (0.02) & 0.995 (0.00) & 0.953 (0.01) \\
& 5  & 0.899 (0.06) & 0.794 (0.02) & 0.947 (0.02) & 0.939 (0.02) & 0.978 (0.01) & 0.821 (0.02) \\
& 10 & 0.857 (0.08) & 0.614 (0.02) & 0.920 (0.02) & 0.924 (0.02) & 0.787 (0.04) & 0.600 (0.02) \\
& 25 & 0.859 (0.06) & 0.292 (0.02) & --            & --            & 0.327 (0.03) & 0.311 (0.02) \\
\midrule
\multirow{5}{*}{3}
& 2  & 0.966 (0.02) & 0.902 (0.01) & 0.983 (0.01) & 0.972 (0.02) & 0.990 (0.01) & 0.912 (0.02) \\
& 3  & 0.941 (0.04) & 0.871 (0.02) & 0.969 (0.01) & 0.952 (0.02) & 0.995 (0.00) & 0.926 (0.01) \\
& 5  & 0.906 (0.06) & 0.788 (0.02) & 0.968 (0.01) & 0.955 (0.02) & 0.980 (0.01) & 0.874 (0.02) \\
& 10 & 0.857 (0.07) & 0.614 (0.02) & 0.803 (0.02) & 0.801 (0.03) & 0.808 (0.04) & 0.606 (0.02) \\
& 25 & 0.857 (0.05) & 0.294 (0.02) & 0.480 (0.03) & 0.483 (0.03) & 0.340 (0.02) & 0.319 (0.02) \\
\bottomrule
\end{tabular}
\end{adjustbox}
\end{table*}

For the benchmark methods (ECOD, IFOREST, LOF, MCD, and OCSVM), we use implementations available in the \texttt{PYOD} package \cite{pyod_package}. For the ECOD, MCD, and OCSVM methods, MAD-based thresholding from the \texttt{pythresh} package \cite{pythresh} is employed, as preliminary visual inspections on single simulations indicated that this approach provides the best performance among the tested thresholding methods in \texttt{pythresh}, while maintaining low computational cost. For the IFOREST and LOF methods, thresholding techniques from \texttt{pythresh} are not consistent; therefore, we apply a custom dynamic thresholding procedure. Using a fixed contamination threshold in these cases would constitute an artificial performance improvement, as such a threshold is not known in real-world data applications.

\textcolor{black}{What is important, during the preliminary evaluation phase, a wide range of thresholding techniques available in the \texttt{pythresh} package was assessed for ECOD, MCD, and OCSVM. As was mentioned above, MAD-based approach for trianing set preparing was ultimately selected because it provided the most accurate results (on smaller subset of simulations) while maintaining low computational overhead. In contrast, the proposed method employs a dedicated filtering procedure integrated into its training stage as part of the developed methodology; this procedure is not based on \texttt{pythresh} and is therefore not directly transferable to the remaining anomaly detection methods, which are consequently compared using the MAD-based filtering scheme best suited to them.}

\begin{figure}[h!]
    \centering
    \includegraphics[width=0.8\linewidth]{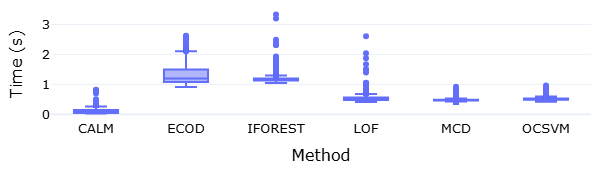}
    \caption{Prediction time comparison of anomaly detection methods for simulated data.}
    \label{fig:times_comp}
\end{figure}

We assume that in a complex system, model fitting (retraining) will occur infrequently. In contrast, label prediction (anomaly or not) will be performed in real time. Fig. \ref{fig:times_comp} presents a comparison of prediction times for CALM and for baseline methods adapted to a real-time use case. As shown, CALM minimizes the latency associated with model computations most effectively.
Tables~\ref{tab:precision_005}--\ref{tab:accuracy_005} summarize the comparative performance of the considered anomaly detection methods across different simulation scenarios and data dimensionalities. Benchmark methods such as MCD and IFOREST achieve very high precision and accuracy in low- and medium-dimensional settings; however, their recall degrades substantially as the dimensionality increases, leading to a significant drop in the F1-score. ECOD exhibits competitive performance only in low-dimensional scenarios, but its effectiveness rapidly deteriorates with increasing dimensionality. LOF shows stable behavior in moderate dimensions, yet its recall and overall robustness decrease in higher-dimensional settings.

In contrast, the proposed CALM method demonstrates consistently balanced performance across all considered metrics and scenarios. In particular, CALM maintains high recall while preserving reasonable precision, resulting in stable F1-scores even in higher-dimensional regimes. This behavior indicates an improved robustness of CALM to the curse of dimensionality compared to the benchmark methods. Moreover, CALM does not rely on prior knowledge of the contamination level, which makes its performance more representative of real-world deployment conditions. Overall, the results indicate that CALM provides a reliable trade-off between detection sensitivity and false-alarm control across a wide range of operating conditions.

\section{Case Study: Unified Test Environment (UTE)}
In this section, we present results obtained from real-world data provided by Nokia. The section includes an analysis of global anomalies detected over a long-term time horizon, with particular consideration of actual failures that occurred during the analyzed period.

\label{sec:real_data}
\subsection{Environment description}

\begin{figure}[h!]
    \centering
    \includegraphics[width=\linewidth]{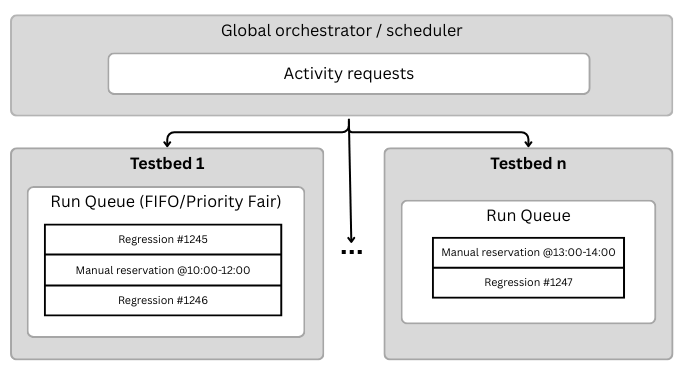}
    \caption{High-level architecture of the testbed scheduling and orchestration system.}
    \label{fig:architecture_scheduler_tedbeds}
\end{figure}

{The analyzed complex testbeds include critical components such as baseband and radio modules, core network elements, transport systems, terminals, and measurement equipment. Additionally, it incorporates supporting infrastructure like servers hosting virtual machines with testing frameworks and necessary software. On top of that network switches, power supplies, and other essential resources.}

{Fig. \ref{fig:architecture_scheduler_tedbeds} illustrates the high-level architecture of the testbed scheduling and orchestration system. A global orchestrator collects activity requests and distributes them across multiple testbeds. Each testbed maintains an independent run queue where tasks are scheduled according to a FIFO or priority-fair policy while also accommodating manually reserved execution windows. This mechanism enables coordinated resource utilization across heterogeneous test environments while ensuring predictable execution of automated and manually scheduled experiments.}

\textcolor{black}{The dataset employed in this study is collected from more than 1{,}000 testbeds. The testbed landscape is inherently dynamic and non-stationary. New configurations are continuously introduced while obsolete ones are progressively decommissioned, mirroring the evolution of the underlying technology and the associated testing requirements. Consequently, the collected data reflects genuine, real-world conditions rather than a static snapshot. Moreover, testbed utilization is governed by continuously evolving software testing cycles: certain configurations are heavily reused, whereas others are only sporadically exercised depending on specific testing needs. This behavior gives rise to a naturally imbalanced usage distribution across the testbed population. Data acquisition is continuous and spans multiple years across geographically distributed laboratories. The system adopts an event-driven paradigm in which each reservation generates a corresponding record. On average, several thousand reservations are logged per day, with hundreds of concurrent reservations sustained in a $24/7$ operating regime. Each record is described by numerical features representing the time durations of key operational phases, namely preparation, waiting, and allocation. The preparation and waiting phases exhibit distributions with durations extending up to several tens of minutes. In contrast, the allocation phase is highly skewed: its median is equal to zero and the majority of values are concentrated near zero, which introduces a strongly imbalanced feature distribution and constitutes one of the principal challenges for the detection task. The input to the pipeline corresponds directly to structured raw time-duration measurements derived from system events. No advanced preprocessing is performed at the data collection stage; instead, any required feature transformation is delegated to the CALM model pipeline, thereby preserving the integrity of the raw measurements and ensuring reproducibility of the ingestion procedure. The analysis is carried out after reservation completion, once all phase durations are known, and is executed periodically (e.g., every 15 minutes). This scheduling strategy yields near-real-time insights while remaining decoupled from the latency of individual reservations. With respect to computational overhead and deployment feasibility, the volume and structure of the data, i.e. integer-valued time measurements amounting to several hundred up to slightly more than one thousand events per day, introduce negligible processing cost. The solution operates alongside the production systems without degrading their performance, thereby demonstrating its suitability for deployment in real-time, production-scale environments.}
 
\subsection{Real data analyses}

The analysis is based on time-related measurements associated with the testbed reservation process. In particular, we focus on two key temporal components that characterize the lifecycle of a reservation request:

\begin{itemize}
    \item \textit{Preparation time} (PT) refers to the interval during which an available testbed is automatically reset and configured to the expected initial state. This phase includes actions such as powering on modules, updating software components, and enabling or disabling required functionalities.
    \item \textit{Waiting time} (WT) is defined as the sum of the allocation time and the \textit{preparation time}, representing the total duration from the submission of a reservation request until the testbed is fully prepared and ready for use.
\end{itemize}

The reservations of individual testbeds are assumed to be independent. Nevertheless, already at the level of a single observation of \textit{preparation time} and \textit{waiting time}, it is possible to determine whether the record is anomalous for a given testbed. 
\textcolor{black}{It is confirmed by the analysis of historical reservation data that did not reveal any statistically significant dependencies between the next observations in this two variables, supporting the independence assumption adopted in this work.}

\textcolor{black}{Furthermore, consecutive reservations are assumed to be independent due to the inherently stochastic nature of the reservation process, which is driven by user-defined requests and scheduling decisions. This characteristic further justifies the application of the CALM procedure to the considered dataset. In addition, the data occasionally contain observations with values close to zero. The CALM framework naturally accommodates such cases by incorporating an additional low-variance noise component into the data model, thereby improving robustness in the presence of near-zero observations.}

In the case of aggregated results across multiple testbeds, a temporal aggregation window must be introduced in order to assess the occurrence of global anomalies. The adopted procedure operates as follows: for each testbed, the online anomaly detection is performed continuously using the procedures described in Algorithms~\ref{alg:train} and~\ref{alg:pred} using the same parameters as in Sec. \ref{sec:simulated_data}. Subsequently, the detected anomalies are aggregated to the global level using Algorithm~\ref{alg:global_ad}.

\begin{figure}[h!]
    \centering
    \includegraphics[width=0.8\linewidth]{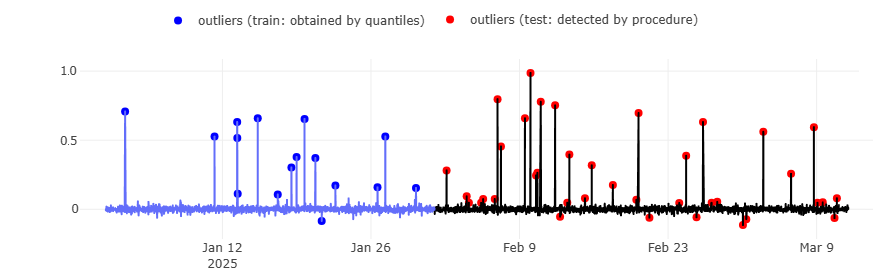}
    \caption{Anomaly identification example for a single testbed (the time values are normalized).}
    \label{fig:single_real}
\end{figure}

In the considered application, a time windows $\omega = 1h$ is used, and the aggregation threshold is set to $T_a = 2\%$. \textcolor{black}{The threshold $T_a$ was determined experimentally based on historical data and (delayed) fault information.} An example of anomaly identification results for a single testbed is presented in Fig.~\ref{fig:single_real}. 

The blue region denotes the training dataset, where anomalies used for training the method are identified using the same procedure as described in Sec.~\ref{sec:simulated_data}. Test data are shown in black, while anomalies detected online are marked with red points.

The obtained results demonstrate a reasonable alignment with the expected behavior of the system. The temporal extent of detected anomalies can be adjusted by appropriately tuning the parameters of the CALM method. The parameter $N_r$ controls the stability of the estimated thresholds by regulating the number of bootstrap resamples, where larger values generally yield more reliable quantile estimates at the cost of increased computational complexity. The quantile level $q_b$ determines the strictness of anomaly acceptance, while the sampling fraction $\phi$ balances bias and variance in the bootstrap approximation by adjusting the effective sample size used for temporary density estimation.

{Importantly, monitoring complex engineering systems increasingly relies on high-frequency signals whose statistical properties vary over time subject to structural changes caused by evolving operating conditions, aging, or faults. Therefore, the model for identifying anomalies should take into account the possibility of retraining. The proposed framework for identification of anomalies can be extended for a two-stage continual learning paradigm. In the first stage, streaming data are segmented online using scalable change-point detection methods. In parallel, anomalies are continuously identified using an unsupervised kernel density estimation approach applied to both aggregated features and raw observations. In the second stage, only statistically significant change points authorize automatic re-estimation of anomaly detection model parameters. This selective adaptation prevents unnecessary retraining, reduces false alarms, and enables continual learning without compromising stability.}

\subsection{Assessment of the framework for global anomaly detection for real-data}
\label{sec:global_anomalies}

Tab.~\ref{tab:waiting_preparation} presents the percentage of detected anomalies for \textit{waiting} and \textit{preparation times} with category, description and root cause of the anomaly that occurs in the analyzed time-window. We use the abbreviation SSP for Support System Problems. The reported value denotes the proportion of testbeds with anomalies that had reservations within a given time window relative to all testbeds that had reservations in that same window (it is important to note that the reference is not the entire population of unused testbeds). From an operational perspective, an additional condition may be introduced indicating the number of testbeds at which the alarm system should trigger. By default, due to the requirement of minimal delays, a one-hour window is used; however, other windows lengths could be used as well. Each row of the table corresponds to a different event described by a unique identifier. The events correspond to $N$ categories listed in the \emph{Problem classification} column. {Within-category analysis reveals distinct patterns in the observed anomaly distribution. Failures classified as \textit{Environment} and \textit{Build/Package} tend to affect a large proportion of testbeds simultaneously, with anomaly rates reaching up to 50-100\% for both waiting and preparation times. This behavior indicates systemic disruptions originating from shared infrastructure components or external dependencies. In contrast, \textit{Interface} anomalies typically affect a smaller but still significant fraction of testbeds (approximately 2--21\% for waiting time and 6-25\% for preparation time), which suggests partial connectivity or integration problems rather than global outages. The \textit{System} category exhibits moderate anomaly levels, reflecting situations where the operational state of the infrastructure introduces delays on a subset of resources. Finally, the \textit{Function} category corresponds to internally caused defects in the orchestration software and is characterized by relatively low anomaly rates (around 3\% and 2\% for waiting and preparation times, respectively), indicating localized software faults rather than widespread infrastructure failures}. As can be observed, for most failures anomalous, behavior is detected in the range of 10-100\% of testbeds for waiting time and 4-100\% for \textit{preparation time}. Setting appropriate thresholds for the alarm system appears to be crucial. For failures affecting a small proportion of testbeds (e.g., ID~12), it may be beneficial to analyze the anomaly rate within a geographic region or topology, which could indicate software-related or mechanical failures.

\begin{table*}[h!]
\centering
\caption{Summary of anomalies and root causes}
\label{tab:waiting_preparation}
\begin{adjustbox}{width=\textwidth}
\begin{tabular}{c c c l c c l}
\hline
\textbf{ID} & \textbf{WT} & \textbf{PT} & \textbf{Description} & \textbf{Classification} & \textbf{Root cause} & \textbf{Suggested follow-up} \\ \hline
1  & 36.36 & 4.55  & Maintenance window in an external system $\rightarrow$ SSP & Environment & external & Monitor / no action \\ 
2  & 20.76 & 25.26 & Connection breakdown with external system $\rightarrow$ I/O & Interface & external & Raise to external system according to SLA \\ 
3  & 2.44  & 6.50  & Connection breakdown with external system $\rightarrow$ I/O & Interface & external & Raise to external system according to SLA \\
4  & 4.47  & 3.91  & Race condition with external system $\rightarrow$ SSP & Environment & external & Raise to external system according to SLA \\
5  & 50.38 & 67.94 & Accidental release of broken external dependency $\rightarrow$ change management / SSP & Build, Package & external & RCA to better understand the cause \\
6  & 46.53 & 64.90 & Accidental release of broken external dependency $\rightarrow$ change management / SSP & Build, Package & external & RCA to better understand the cause \\
7  & 11.46 & 4.17  & System state at the time of anomaly caused more delays than usual & System & external & Monitor / no action \\
8  & 12.58 & 10.69 & Connection with external system $\rightarrow$ I/O & Interface & external & Raise to external system according to SLA \\
9  & 16.30 & 18.58 & Connection with external system $\rightarrow$ I/O & Interface & external & Raise to external system according to SLA \\
10 & 100.00 & 100.00 & Maintenance window in an external system $\rightarrow$ SSP & Environment & external & Monitor / no action \\
11 & 100.00 & 25.00 & Maintenance window in an external system $\rightarrow$ SSP & Environment & external & Monitor / no action \\
12 & 2.97  & 2.13  & Error in the orchestrator software $\rightarrow$ software logic & Function & internal & Escape Defect Analysis \\ \hline
\end{tabular}
\end{adjustbox}
\end{table*}

\section{Conclusion \textcolor{black}{and further works}}
\label{sec:summary}
In this work, we introduced CALM, an unsupervised anomaly detection method designed for reliable operation in high-dimensional and realistic settings.  
CALM achieves rapid detection, identifying anomalies in real-time.  
The method is inherently flexible, offering multiple parameterization options that allow practitioners to balance detection sensitivity and false-alarm rates according to operational needs.  
Importantly, CALM does not require labeled data or prior knowledge of the contamination level, which makes it well suited for real-world deployments.  
Benchmark methods such as MCD, IFOREST, ECOD, and LOF show strong performance in low- and medium-dimensional scenarios, but their recall and F1-scores degrade significantly as dimensionality increases.  This degradation highlights their sensitivity to the curse of dimensionality and limits their robustness in complex environments.  
In contrast, CALM demonstrates consistently balanced performance across all evaluated scenarios and metrics.  In particular, it maintains high recall while preserving reasonable precision, resulting in stable F1-scores even in high-dimensional regimes.  
These results indicate that CALM provides improved robustness compared to existing benchmarks. Overall, CALM offers a reliable and practical trade-off between fast detection, accuracy, and robustness, making it a strong candidate for deployment in diverse operational conditions. {While the proposed methodology is validated within a specific telecommunication framework, it is highly transferable to numerous application domains where anomaly detection is a critical issue. One of the example is condition monitoring, where the identification of anomalies in diagnostic signals may indicate damage of the machine or measurement error. Providing near-instantaneous feedback in such contexts can significantly enhance the operational efficiency of monitoring system.}

\textcolor{black}{The proposed framework has been specifically designed for independent data streams, which are common in the considered reservation-monitoring scenario. Future work will focus on extending the methodology to more complex settings, including data exhibiting temporal periodicity, without requiring extensive preprocessing or explicit seasonality removal. Although we have previous experience with multivariate statistical procedures, including hypothesis testing based on multivariate CDFs and PDFs \cite{WITULSKA2026103445}, such approaches become computationally demanding in higher dimensions and typically require multiple observations before an anomaly can be reliably identified. This significantly limits their applicability in continuous monitoring systems, where low latency and rapid detection are critical. A promising direction is to incorporate inter-feature dependencies through dedicated feature engineering, e.g., by constructing joint descriptors or statistics derived from multiple variables. The effectiveness of such dependency-aware extensions, however, remains application-dependent and will be investigated in future studies.}

\textcolor{black}{The bootstrap threshold ($q_b$) controls the nominal false positive rate only at the marginal level. In general, the overall false positive rate may increase with dimensionality. If this problem occur, the possible extension is to introduce an additional score aggregation mechanism, allowing low-significance isolated anomalies to be filtered out while emphasizing observations that are repeatedly identified as anomalous across multiple dimensions. Such aggregation-based thresholding may provide better control of the global false positive rate in higher-dimensional applications and will be investigated in future work.}

\section*{Acknowledgements}
    The work of JW and AW is supported by the NCN Weave-Unisono project entitled “Advanced signal processing techniques for cyclostationary modelling in Gaussian and non-Gaussian noisy environment -detection of cyclic sources, estimation, optimisation of algorithms and validation in the context of fault identification” (No. \\ 2025/07/Y/ST8/00070).

\section*{Conflict of interest}
No conflicts of interests.

\appendix
\section*{Appendix}
\textcolor{black}{Here, we present the comparative evaluation of using $s(x;\Theta,j)=- \sqrt{(exp(f^{\Theta}_{i,j}(x))}$ and \\ $s(x;\Theta,j)= -log{(exp(f^{\Theta}_{i,j}(x))} = -f^{\Theta}_{i,j}(x)$ as a scoring functions in CALM algorithm.}

\textcolor{black}{The same simulations as those described in Section~III were performed for dimensions $d = 2, 3,$ and $5$ (for visualization purposes). The CALM method was evaluated using both scoring functions, and the resulting values of the performance metrics---precision, recall, F1-score, and accuracy---are summarized in Tables~\ref{precision_scoring_comp}--\ref{accuracy_scoring_comp}. For each dimension and each scenario 100 Monte Carlo trials were made.} \textcolor{black}{As can be observed, the values of all evaluation metrics obtained with the two scoring functions are very similar for simulation scenarios 2 and 3. For simulation scenario 1, however, the original scoring function based on the square root yields higher Recall and F1-score values than the logarithmic scoring function.}

\begin{table*}[!t]
\centering
\begin{minipage}[t]{0.48\textwidth}
\centering
\caption{Precision for scoring functions (mean and standard deviation).}
\label{tab:precision_scoring_functions}
\begin{adjustbox}{width=0.6\textwidth}
\begin{tabular}{cccc}
\toprule
\textbf{Sim.} & \textbf{Dim.} & \textbf{Logarithmic} & \textbf{Sqrt} \\
\midrule
\multirow{3}{*}{1}
& 2 & 0.968 (0.04) & 0.965 (0.04) \\
& 3 & 0.973 (0.02) & 0.970 (0.03) \\
& 5 & 0.975 (0.02) & 0.970 (0.02) \\
\midrule
\multirow{3}{*}{2}
& 2 & 0.964 (0.04) & 0.976 (0.03) \\
& 3 & 0.966 (0.03) & 0.966 (0.02) \\
& 5 & 0.970 (0.03) & 0.969 (0.02) \\
\midrule
\multirow{3}{*}{3}
& 2 & 0.973 (0.04) & 0.974 (0.03) \\
& 3 & 0.964 (0.03) & 0.966 (0.03) \\
& 5 & 0.974 (0.02) & 0.969 (0.02) \\
\bottomrule
\label{precision_scoring_comp}
\end{tabular}
\end{adjustbox}
\end{minipage}
\hfill
\begin{minipage}[t]{0.48\textwidth}
\centering
\caption{Recall for scoring functions (mean and standard deviation).}
\label{tab:recall_scoring_functions}
\begin{adjustbox}{width=0.6\textwidth}
\begin{tabular}{cccc}
\toprule
\textbf{Sim.} & \textbf{Dim.} & \textbf{Logarithmic} & \textbf{Sqrt} \\
\midrule
\multirow{3}{*}{1}
& 2 & 0.614 (0.13) & 0.718 (0.14) \\
& 3 & 0.620 (0.11) & 0.702 (0.09) \\
& 5 & 0.647 (0.10) & 0.712 (0.09) \\
\midrule
\multirow{3}{*}{2}
& 2 & 0.702 (0.15) & 0.703 (0.13) \\
& 3 & 0.706 (0.11) & 0.704 (0.13) \\
& 5 & 0.724 (0.09) & 0.716 (0.09) \\
\midrule
\multirow{3}{*}{3}
& 2 & 0.713 (0.13) & 0.698 (0.13) \\
& 3 & 0.691 (0.10) & 0.720 (0.11) \\
& 5 & 0.729 (0.09) & 0.726 (0.08) \\
\bottomrule
\end{tabular}
\end{adjustbox}
\end{minipage}
\end{table*}

\begin{table*}[!t]
\centering
\begin{minipage}[t]{0.48\textwidth}
\centering
\caption{F1-score for scoring functions (mean and standard deviation).}
\label{tab:f1_scoring_functions}
\begin{adjustbox}{width=0.6\textwidth}
\begin{tabular}{cccc}
\toprule
\textbf{Sim.} & \textbf{Dim.} & \textbf{Logarithmic} & \textbf{Sqrt} \\
\midrule
\multirow{3}{*}{1}
& 2 & 0.743 (0.10) & 0.815 (0.09) \\
& 3 & 0.751 (0.08) & 0.810 (0.07) \\
& 5 & 0.774 (0.07) & 0.818 (0.06) \\
\midrule
\multirow{3}{*}{2}
& 2 & 0.803 (0.10) & 0.810 (0.10) \\
& 3 & 0.811 (0.08) & 0.808 (0.09) \\
& 5 & 0.826 (0.06) & 0.820 (0.06) \\
\midrule
\multirow{3}{*}{3}
& 2 & 0.816 (0.09) & 0.806 (0.09) \\
& 3 & 0.801 (0.07) & 0.820 (0.07) \\
& 5 & 0.830 (0.06) & 0.827 (0.05) \\
\bottomrule
\end{tabular}
\end{adjustbox}
\end{minipage}
\hfill
\begin{minipage}[t]{0.48\textwidth}
\centering
\caption{Accuracy for scoring functions (mean and standard deviation).}
\label{tab:accuracy_scoring_functions}
\begin{adjustbox}{width=0.6\textwidth}
\begin{tabular}{cccc}
\toprule
\textbf{Sim.} & \textbf{Dim.} & \textbf{Logarithmic} & \textbf{Sqrt} \\
\midrule
\multirow{3}{*}{1}
& 2 & 0.960 (0.01) & 0.970 (0.01) \\
& 3 & 0.943 (0.02) & 0.955 (0.01) \\
& 5 & 0.916 (0.02) & 0.930 (0.02) \\
\midrule
\multirow{3}{*}{2}
& 2 & 0.968 (0.01) & 0.970 (0.01) \\
& 3 & 0.954 (0.02) & 0.953 (0.02) \\
& 5 & 0.933 (0.02) & 0.930 (0.02) \\
\midrule
\multirow{3}{*}{3}
& 2 & 0.970 (0.01) & 0.969 (0.01) \\
& 3 & 0.952 (0.02) & 0.957 (0.02) \\
& 5 & 0.933 (0.02) & 0.932 (0.02) \\
\bottomrule
\end{tabular}
\label{accuracy_scoring_comp}
\end{adjustbox}
\end{minipage}
\end{table*}

\bibliographystyle{elsarticle-num} 
\bibliography{mybibliography}

\end{document}